\documentstyle[12pt]{article}
\def\ba{\begin{eqnarray}}
\def\be{\begin{equation}}
\def\ea{\end{eqnarray}}
\def\ee{\end{equation}}
\def\ft#1#2{{\textstyle {{\scriptstyle #1} \over {\scriptstyle #2}}}}

\def\ww3{{$W_3$}}

\def\del{\partial}
\def\p{\partial}

\begin{document}
\topmargin 0pt
\oddsidemargin 5mm
\begin{titlepage}
\begin{flushright}
CTP TAMU-9/95\\
hep-th/9504122\\
\end{flushright}
\vspace{1.0truecm}
\begin{center}
{\bf {\Large A Search for New $(2,2)$ Strings}}
\vspace{1.5truecm}

{\large H. L\"u\footnote{Supported in part by the
U.S. Department of Energy, under grant DE-FG05-91-ER40633,\hfil\break
\phantom{E-m}E-mail: hong@phys.tamu.edu, pope@phys.tamu.edu.},\ \  C.N.
Pope$^{1}$, and  E.
Sezgin\footnote{Supported in part by the National Science Foundation, under
grant PHY-9411543,\hfil\break
\phantom{E-m}E-mail: sezgin@phys.tamu.edu.}}
\vspace{1.1truecm}

{\small Center for Theoretical Physics, Texas A\&M University,
                College Station, TX 77843-4242} \vspace{1.1truecm}


\end{center}
\vspace{0.5truecm}

\begin{abstract}
\vspace{1.0truecm}
There are at present two known string theories in $(2,2)$ dimensions.
One of them is the well known $N=2$ string, and the other one is a more
recently constructed $N=1$ spacetime supersymmetric string. They are both
based on certain twistings and/or truncations of the small $N=4$
superconformal algebra, realised in terms of $(2,2)$ superspace variables.
In this paper, we investigate more general possibilities for string theories
based on algebras built with the same set of fields. We find that there
exists one more string theory, based on an algebra which is not contained
within the $N=4$ superconformal algebra. We investigate the spectrum and
interactions of this theory.

\end{abstract}
\end{titlepage}
\newpage
\pagestyle{plain}
\section{Introduction}

There are at present two known string theories in 2+2 dimensions. One of
them is based on the $N=2$ superconformal algebra \cite{foot}, and it has
the interesting feature of describing  self-dual gravity in 2+2 dimensions
\cite{ov}. One of the surprising features of this model is that
notwithstanding the worldsheet supersymmetry, it lacks spacetime
supersymmetry \cite{ov}. In search of an alternative string model in 2+2
dimensions which does exhibit spacetime supersymmetry, a second string model
was recently proposed in \cite{klpswx}. The underlying conformal algebra in
this case is a twisted and truncated version of the ``small'' $N=4$
superconformal algebra. This model indeed has spacetime supersymmetry;
however, it turns out that the physical spectrum contains an infinite tower
of massive states, and the massless states do not describe supergravity in
2+2 dimensions \cite{klpswx}.

Both of the above models can be constructed by making use of bilinear
combinations of the bosonic coordinates $X^{\alpha\dot\alpha}$, fermionic
coordinates $\theta^\alpha$, and their conjugate momenta $p_\alpha$, to
build the currents of the underlying  worldsheet algebras. The indices
$\alpha$ and ${\dot \alpha}$ label the two-dimensional spinor
representations of $SL(2)_R\times SL(2)_L \approx SO(2,2)$. Note that because
the Lorentz group factorises, we can use spinors of one handedness only.

In terms of these variables, the currents for the $N=2$ superconformal
algebra are
\begin{eqnarray}
T&=& -\ft12 \del X^{\alpha\dot\alpha}\del X_{\alpha\dot\alpha}
     -p_\alpha \del \theta^\alpha \ , \qquad J=p_{\alpha}\theta^{\alpha}\ ,
\nonumber\\
G^{\dot 1} &=& \theta_{\alpha}\del X^{\alpha\dot 1}\ , \qquad
G^{\dot 2} = p_{\alpha}\del X^{\alpha\dot 2} \ .  \label{n2alg}
\end{eqnarray}
This is a twisted version of the usual realisation, since here
the $(p,\theta)$ system has dimension $(1,0)$.  In this realisation,
manifest $SL(2)_L$ invariance is broken. It can be restored \cite{ws1}
by enlarging the
set of currents to those of the small $N=4$ superconformal algebra,
given, again in a twisted basis, by
\begin{eqnarray}
T&=& -\ft12 \del X^{\alpha\dot\alpha}\del X_{\alpha\dot\alpha}
     -p_\alpha \del \theta^\alpha \ , \nonumber\\
G^{\dot \alpha} &=& p_{\alpha}\del X^{\alpha\dot \alpha}\ , \qquad
{\widetilde G}^{\dot \alpha} = \theta_{\alpha}\del X^{\alpha\dot \alpha} \ ,
\label{n4alg}\\
 J_0 &=& p_{\alpha}\theta^{\alpha}\ ,
\qquad J_+= p_\alpha p^\alpha\ , \qquad J_-=\theta_\alpha\theta^\alpha\ .
\nonumber
\end{eqnarray}
Naively, this system appears to be non-critical. However, the
currents are reducible, and a proper quantisation requires the
identification of the irreducible subset \cite{ws1}, namely the $N=2$ currents
given in (\ref{n2alg}).

The currents of the second string model discussed above, which has $N=1$
spacetime supersymmetry, can also be expressed in a fully covariant way.
They are given by \cite{klpswx}
\begin{eqnarray}
T&=& -\ft12 \del X^{\alpha\dot\alpha}\del X_{\alpha\dot\alpha}
     -p_\alpha \del \theta^\alpha \ , \qquad
 G^{\dot \alpha} = p_{\alpha}\del X^{\alpha\dot\alpha} \ ,
\qquad J = p_\alpha p^\alpha \ .
\label{tn4alg}
\end{eqnarray}
In fact, these currents are a truncation of the realisation of the small
$N=4$ superconformal algebra given in (\ref{n4alg}). All the currents have
spin two, and hence the system is critical, with zero central charge.
However, this set of currents is reducible. An irreducible subset, which is
also critical, is given by $T$ and $G^{\dot 1}$. The BRST cohomology of this
system was studied in Ref.\ \cite{klpswx}.

It is interesting to investigate if there can exist other string theories in
2+2 dimensions, based on other bilinear currents built from the
same ingredients. In this paper, we investigate this question systematically,
and we show that indeed there exists one more possibility. The criteria that
lead to this conclusion are, in addition to closure, that the currents are
irreducible, and that they give rise to a critical realisation of the
worldsheet algebra. The set of currents we have found is
\begin{eqnarray}
T&=& -\ft12 \del X^{\alpha\dot\alpha}\del X_{\alpha\dot\alpha}
     -p_\alpha \del \theta^\alpha \ , \qquad J =\del (\theta_\alpha
\theta^\alpha) \ , \nonumber\\
 G^{\dot 1} &=& p_{\alpha}\del X^{\alpha\dot 1} \ , \qquad
{\widetilde G}^{\dot 1}=\theta_\alpha \del X^{\alpha\dot 1}\ .
\label{newalg}
\end{eqnarray}
It is easy to see that the currents $(T, G^{\dot 1},
{\widetilde G}^{\dot 1}, J)$ have spins $(2,2,1,1)$.
In addition to the standard OPEs of $T$ with $(T,J,G^{\dot 1},{\widetilde
G}^{\dot 1})$, the only non-vanishing OPE is
\begin{equation}
J(z)\, G^{\dot 1} (w) \sim {2{\widetilde G}^{\dot 1}\over (z-w)^2}+
{\del {\widetilde G}^{\dot 1}\over (z-w)} \ .
\end{equation}
This algebra is related to the small $N=4$ superconformal algebra, not
directly as a subalgebra, but in the following way. The subset of currents
$T, G^{\dot 1}, \widetilde G^{\dot 1}$ and $J_-$ in (\ref{n4alg}) form a
critical closed algebra. However these currents form a reducible set. To
achieve irreducibility, we simply differentiate the current $J_-$, thereby
obtaining the set of currents given in (\ref{newalg}). Note that taking the
derivative of $J_-$ still gives a primary current with the same anomaly
contribution, since $12s^2-12s+2$ takes the same value for $s=0$ and $s=1$.
We shall study the string theory based on this
algebra.\footnote{We have concentrated on conformal algebras which make use
of the $SL(2)_R$ spinor variables $(p_\alpha, \theta^\alpha)$.
{\it A priori}, we could also use the $SL(2)_L$ spinor variables  $(p_{\dot
\alpha}, \theta^{\dot \alpha})$. Interestingly enough, no new critical
algebra seems to arise in this fashion.}

     To proceed with the BRST quantisation of the model, we introduce the
fermionic ghost fields $(c,b)$ and $(\gamma,\beta)$ for the currents $T$ and
$J$, and the bosonic ghost fields $(s,r)$ and $({\tilde s},{\tilde r})$ for
$G^{\dot 1}$ and ${\widetilde G}^{\dot 1}$. It is necessary to bosonise the
commuting ghosts, by writing $s=\eta e^\phi$, $r=\del \xi e^{-\phi}$,
$\tilde s=\tilde \eta e^{\tilde \phi}$ and $\tilde r=\del \tilde\xi
e^{-\tilde\phi}$. The BRST operator for the model is then given by
\begin{eqnarray}
Q&=& \oint c\Big( -\ft12 \del X_{\alpha\dot\alpha}\del X^{\alpha\dot\alpha}
-p_\alpha \del \theta^\alpha-\ft12 ( \del \phi)^2 -\ft12 (\del\tilde \phi)^2
-\ft32 \del^2 \phi -\ft12 \del^2 \tilde\phi \nonumber \\
&&  -\eta\del\xi -\tilde\eta \del \tilde \xi -b\del c
-\beta\del\gamma\Big) \label{brst} \\
&&+\eta e^\phi p_\alpha \del X^{\alpha \dot 1} +
\tilde \eta e^{\tilde\phi} \theta_\alpha \del X^{\alpha \dot 1} +
\del \gamma \Big( \ft12 \theta^\alpha\theta_\alpha -\del\tilde\xi\eta
e^{\phi-\tilde\phi}\Big)\ .  \nonumber
\end{eqnarray}

Since the zero modes of $\xi$ and $\tilde\xi$ do not appear in the BRST
operator, there exist BRST non-trivial picture-changing operators:
\ba
Z_\xi &=& \{ Q, \xi\}=c\del\xi+e^\phi p_\alpha\del X^{\alpha\dot 1}
-\del \gamma \del\tilde\xi e^{\phi-\tilde\phi}\ , \nonumber\\
Z_{\tilde\xi} &=& \{ Q, \tilde\xi\}=c\del\tilde\xi + e^{\tilde\phi}
\theta_\alpha\del X^{\alpha\dot 1}\ . \label{pic}
\ea
It turns out that these two picture changers are not invertible. Thus, one has
the option of including the zero modes of $\xi$ and $\tilde\xi$ in the Hilbert
space of physical states. This would not be true for a case where the picture
changers were invertible. Under these circumstances, the inclusion of the
zero modes would mean that all physical states would become trivial, since
$|{\rm phys}\rangle =Q(\xi Z_\xi^{-1} |{\rm phys}\rangle)$. However, in this
paper we shall nevertheless choose to exclude the zero modes of $\xi$ and
$\tilde\xi$ from the Hilbert space. It is interesting to note that in this
model the zero mode of the ghost field $\gamma$ for the spin--1 current is
also absent in the BRST operator. If one excludes this zero mode from
the Hilbert space, one can then introduce the corresponding picture-changing
operator $Z_\gamma = \{Q,\gamma\}=c\del \gamma$. In this paper, we shall
indeed choose to exclude the zero mode of $\gamma$.

In order to discuss the cohomology of the BRST operator (\ref{brst}), it is
convenient first to define an inner product in the Hilbert space. Since the
zero modes of the $\xi,\tilde\xi$ and $\gamma$ are excluded, the inner
product is given by
\be
\langle \del^2 c\,\del c\, c\, \theta^\alpha\theta_\alpha\,
e^{-3\phi-\tilde\phi}\rangle=1\ . \label{innerp}
\ee

Let us first discuss the spectrum of massless states in the Neveu-Schwarz
sector. The simplest such state is given by
\be
 V=c\, e^{-\phi-\tilde\phi} e^{ip\cdot X} \ . \label{massless}
\ee
As in the case of the $N=2$ string discussed in \cite{lp}, since the
picture-changing operators are not invertible the massless states in
different pictures cannot necessarily all be identified. In fact, the
picture changers annihilate the massless operators such as (\ref{massless})
when the momentum $p^{\alpha\dot 1}$ is zero. However, massless operators in
other pictures still exist at momentum $p^{\alpha\dot 1}=0$. For example, in
the same picture as the physical operator $Z_{\tilde \xi} V$ that vanishes
at $p^{\alpha\dot 1}=0$ is a physical operator that is non-vanishing for all
on-shell momenta, namely
\be
\Psi=h_\alpha\, c\,\theta^\alpha\, e^{-\phi}\, e^{ip\cdot X}\ , \label{psi}
\ee
which is physical provided that $p^{\alpha\dot 1}\, h_\alpha=0$ and
$p_{\alpha\dot\alpha} p^{\alpha\dot\alpha}=0$. In fact, $Z_{\tilde \xi} V$
is nothing but $\Psi$ with the polarisation condition solved by writing
$h^\alpha=p^{\alpha\dot 1}$. However, we can
choose instead to solve the polarisation condition by writing
$h^\alpha=p^{\alpha\dot 2}$, which is non-vanishing even when
$p^{\alpha\dot 1}=0$. Thus, the operators $V$ and $\Psi$ cannot be
identified under picture changing when $p^{\alpha\dot1}=0$. In fact when
$p^{\alpha\dot1}=0$ there is another independent solution for $\Psi$, since
the polarisation condition becomes empty in this case.  A convenient way to
describe the physical states is in terms of $\Psi$ given in (\ref{psi}),
with the polarisation condition re-written in the covariant form
$p^{\alpha\dot\alpha}\, h_\alpha=0$, together with a further physical
operator which is defined only when $p^{\alpha\dot1}=0$.  In this
description, the physical operator $\Psi$ is defined for all on-shell
momenta.

     If one adopts the traditional viewpoint that physical operators related
by picture changers describe the same physical degree of freedom, one would
then interpret the spectrum as containing a massless operator
(\ref{massless}), together with an infinite number of massless operators
that are subject to the further constraint $p^{\alpha\dot1}=0$ on the
on-shell momentum.\footnote{It should be emphasised that the possibility of
having $p^{\alpha\dot 1}=0$ while $p^{\alpha\dot 2}\ne 0$ is a consequence
of our having chosen a real structure on the $(2,2)$ spacetime
\cite{klpswx,lp}, rather than the more customary complex structure
\cite{ov}.} This viewpoint is not altogether satisfactory in a case such as
ours, where the picture changing operators are not invertible. An
alternative, and moreover covariant, viewpoint is that the physical
operators in different pictures, such as $V$ and $\Psi$, should be viewed as
independent. At first sight one might think that this description leads to
an infinite number of massless operators. However, as we shall see later,
the interactions of all the physical operators can be effectively described
by the interaction of just the two operators $V$ and $\Psi$. Thus the theory
effectively reduces to one with just two massless operators, one a scalar
and the other a spinorial bosonic operator.

     Let us now turn our attention to the spectrum of massive states. We find
that there exist no states with negative ${(\rm mass)}^2$. There are states
of positive ${(\rm mass)}^2$ of the following form
\ba
V_n^{(\delta)} &=& c\, (\del^n p)^2\cdots p^2\, e^{n\phi-(n+2)\tilde\phi}\,
                   \p^{2n+2+\delta}\gamma\cdots \p\gamma\, e^{ip\cdot X}\ ,
\nonumber\\
U_n^{(\delta)} &=& c\, (\del^{n+1} \theta)^2\cdots \theta^2\,
   e^{-(n+3)\phi+(n+1)\tilde\phi}\,
                   \p^{2n+1+\delta}\beta\cdots \beta\, e^{ip\cdot X}\ ,
\label{massive}
\ea
where $n\ge -1$, $\delta=0$ or 1 and the mass is given by
${\cal M}^2=(2n+2+\delta) (2n+3+\delta)$. Note that $V_{-1}^{(0)}$
corresponds to the massless operator (\ref{massless}), and
$\p c\, \theta^\alpha\theta_\alpha\, U_{-1}^{(0)}$ is its conjugate. All the
other operators in (\ref{massive}) have ${\cal M}^2 > 0$ and they are in the
non-standard ghost sector.  It is worth remarking that the massive operators
(\ref{massive}) assume a particularly simple form if we bosonise the fields
$(p_\alpha, \theta^\alpha)$ as $p_\alpha=e^{-i\sigma_\alpha}$ and
$\theta^\alpha=e^{i\sigma_\alpha}$. They become purely exponential vertex
operators:
\ba
V_n^{(\delta)} &=& c\, e^{-i(n+1)(\sigma_1+\sigma_2)}e^{n\phi-(n+2)\tilde\phi}
                   \p^{2n+2+\delta}\gamma\cdots \p \gamma\, e^{ip\cdot X}\ ,
\nonumber\\
U_n^{(\delta)} &=& c\, e^{i(n+2)(\sigma_1+\sigma_2)}
   e^{-(n+3)\phi+(n+1)\tilde\phi}
                   \p^{2n+1+\delta}\beta\cdots \beta\, e^{ip\cdot X}\ .
\label{vertex}
\ea
Having written the physical operators in this bosonised form, we can now
allow $n$ to take half-integer as well as integer values, corresponding to
physical operators in Ramond and Neveu-Schwarz sectors respectively.  One
can easily verify that all the NS and R states are local to each other, in
the sense that their OPEs with each other have integer poles.  It follows
from the fact that $n\ge -1$ that there is no massless operator in the R
sector; however, at each massive level, there is one NS operator and one R
operator.

     The restriction that $n$ can only take integer and half-integer values
comes from the fact that we have chosen not to include the zero mode of
the $\gamma$ field in the Hilbert space of physical states.
If we had instead included this zero mode, we could then have bosonised the
$(\gamma, \beta)$ system, in which case it would be $\gamma
V_n^{(\delta)}$ and $U_n^{(\delta)}$ that were physical, for all values of
$n$.   It follows from the mass formula ${\cal M}^2=(2n+2 + \delta)
(2n+3+\delta)$ that the mass spectrum of the physical states would then be
continuous.  Thus the exclusion of the zero mode of $\gamma$ from the
Hilbert space is clearly desirable.

     The r\^ole of the picture-changing operators is quite different for the
massive physical operators from that for the massless operators.  As we
discussed above,  the massless physical operators reside only in the NS
sector.  The picture-changing operators cannot be used to
identify all these massless physical operators in different pictures. Instead,
the theory effectively contains two massless operators.  However, for the
massive physical operators, the picture-changing operators {\it can} be used
to identify the physical operators in different pictures and hence at each
mass level, there is one NS operator and one R operator.  We shall
demonstrate this with an example.  Let us consider the simple massive
operator $V_0^{(0)}$ given in (\ref{massive}).  The physical operator with
the same picture as $Z_{\tilde \xi} V_0^{(0)}$ is given by $\widetilde\Psi =
h^{\alpha}\, c\, p_{\alpha} e^{-\tilde\phi} \del^2\gamma\, \del\gamma\,
e^{ip\cdot X}$.  It is physical provided that $p_{\alpha\dot\alpha}\,
p^{\alpha\dot\alpha} = -6$ and $p^{\alpha\dot1}h_\alpha = 0$.   The
solution for the polarisation spinor is $h^a = p^{\alpha\dot1}$.  Note that
$p^{\alpha\dot1}$ can never be zero for a massive operator, and furthermore
it is not proportional to $p^{\alpha\dot2}$.   It follows that the physical
operator $\widetilde \Psi$ is precisely related to the operator $V^{(1)}_0$
by picture changing, namely $\widetilde \Psi = Z_{\tilde\xi} V^{(1)}_0$.
This example illustrates that the massive physical operators in different
pictures can be identified by picture changers for all on-shell momenta.
Thus, to summarise, at each mass level there are two physical degrees of
freedom.  At the massless level, they correspond to two NS operators; At
each massive level, they correspond to one NS operator and one R operator.

These operators are only a subset of the totality of massive operators
in the theory. In particular, we observe that ${\cal M}^2$ is proportional
to $n^2$. One normally expects that the massive states should have
${\cal M}^2$ proportional to $n$. Indeed, as we shall show later, such
states do occur. However, these types of massive states are more complicated.

We now turn to the discussion of the interactions in the theory.   We first
consider interactions involving only the massless operators.  If we adopt
the interpretation that the massless spectrum is described by the operator
(\ref{massless}), together with the additional operators subject to the
further momentum constraint $p^{\alpha\dot1}=0$, there is a non-vanishing
three-point amplitude given by
\be
\Big \langle Z_{\tilde\xi} V(z_1,p_{(1)})\, Z_{\tilde\xi} V(z_2,p_{(2)})\,
V(z_3,p_{(3)})\Big\rangle= p_{(1)\alpha}^{\dot 1}\, p_{(2)}^{\alpha\dot 1}\ .
\label{3pf1}
\ee
The four-point amplitude $\langle Z_{\tilde\xi} V\,
Z_{\tilde\xi} V\, Z_{\tilde\xi} V\, Z_\xi V\rangle$ vanishes for kinematical
reasons \cite{lp}. We expect that the higher-point amplitudes also vanish for
the same reason.

     The above description is not satisfactory since the result for the
three-point amplitude breaks Lorentz invariance.  Furthermore, it is not
complete since there are an infinite number of physical operators with the
further momentum constraint $p^{\alpha\dot1}=0$ that can interact with each
other.  As we discussed earlier, the massless spectrum can be better
described by the scalar operator (\ref{massless}) together with the spinorial
bosonic operator (\ref{psi}), without the use of the picture-changing
operators.  There is one three-point interaction between these two
operators, namely
\be
\Big \langle \Psi(z_1, p_{(1)})\, \Psi(z_2, p_{(2)})\, V(z_3, p_{(3)})
\Big\rangle = h_{(1)\alpha}\, h_{(2)}^\alpha \ .\label{3pf2}
\ee
Note that this three-point amplitude is manifestly Lorentz invariant, and
reduces to (\ref{3pf1}) if one solves for the polarisation spinors by
writing $h^\alpha = p^{\alpha\dot1}$.   There are also an infinite number of
massless physical operators with different pictures in the spectrum, and
they can all be expressed in a covariant way.  As one steps through the
picture numbers, the character of the physical operators alternates between
scalar and spinorial.  The three-point interactions of all these operators
lead only to the one amplitude given by (\ref{3pf2}).  In view of their
equivalent interactions, all the scalar operators can be identified and all
the spinorial operators can be identified.   The massless spectrum can thus
be effectively described by the scalar operator (\ref{massless}) and the
spinorial operator (\ref{psi}).  All four-point and higher amplitudes
vanish.

   Let us now consider the interactions of the massive states. In particular
we are interested  in four-point amplitudes since they give information
about new massive states.  We find a non-vanishing four-point amplitude,
for physical operators with (mass)$^2$ equal to $0, 0, 6, 6$ respectively,
given by
\ba
&&\Big\langle \Phi\, \Psi\,
 Z_{\tilde\xi} V_0^{(0)}\, Z_\xi U_0^{(0)}\Big\rangle=
 \Big\langle c\, h_{(1)}^\alpha\, p_\alpha\, e^{-\tilde\phi} e^{ip_{(1)}
\cdot X}\,\, c\, h_{(2)}^{\alpha}\,\theta_\alpha\, e^{-\phi} e^{ip_{(2)}
\cdot X} \times\nonumber\\
&&\qquad\qquad \times \oint p_{(3)}^{\alpha\dot 1}\, p_\alpha\, e^{-\tilde\phi}
\p^2\gamma\p\gamma\, e^{ip_{(3)} \cdot X}\,\,
c\, p_{(4)}^{\alpha\dot 1}\, \p \theta_\alpha\, \theta^2\,
e^{-2\phi+\tilde\phi} \p\beta \beta\,  e^{ip_{(4)}\cdot X}\Big\rangle
\nonumber\\
&& =\Big((\ft{u}2 -3 ) A -\ft{s}2 B\Big)
{\Gamma(-\ft{s}2)\, \Gamma(-\ft{t}2 +3) \over \Gamma(\ft{u}2-2)}\ ,
\label{4pf}
\ea
where $s, t$ and $u$ are the Mandlestam variables satisfying $s+t+u=12$,
and  $A=h_{(1)}^\alpha\,
h_{(2)\alpha}\, p_{(3)}^{\beta\dot1}\, p_{(4)\beta}{}^{\dot1}$ and
$B=h_{(1)\alpha}\, p_{(3)}^{\alpha\dot1}\, h_{(2)}^{\beta}\,
p_{(4)\beta}{}^{\dot1}$.   The kinematic factor in the four-point amplitude
(\ref{4pf}) is non-zero in general, implying the existence of an infinite
tower of physical operators whose (mass)$^2$ is linearly dependent on the
integer $n$.  In particular, the sequence of poles in the $s$ channel of the
four-point amplitude implies that there is an infinite tower of massive
physical operators with standard ghost structure.

      Although the theory contains an infinite tower of physical operators,
the massless sector and its interactions are remarkably simple.   In
particular, although all the massive physical operators break Lorentz
invariance, the massless operators and their interactions have manifest
spacetime Lorentz invariance.  If we associate spacetime fields $\phi$ and
$\psi_\alpha$ with the physical operators $V$ and $\Psi$, it follows from
the three-point amplitude (\ref{3pf2}) that we can write the
field equations:
\be
\del_{\alpha\dot\alpha}\del^{\alpha\dot\alpha}\phi=\psi^\alpha\,\psi_\alpha\ ,
\qquad \del_{\alpha\dot\alpha}\psi^\alpha = \psi^\alpha\del_{\alpha\dot
\alpha}\phi\ .\label{feom}
\ee
We have suppressed Chan-Paton group theory factors that must be introduced
in order for the three-point amplitude to be non-vanishing in the open
string.    It is easy to see even from the kinetic terms in the field
equations (\ref{feom}) that there is no associated Lagrangian.  Note that
there is no undifferentiated $\phi$ field, owing to the fact that the theory
is invariant under the transformation $\phi\longrightarrow \phi + {\rm
const.}$   This can be easily seen from the three-point function
(\ref{3pf2}), which vanishes when $p^{\alpha\dot\alpha}_{(3)}=0$,
corresponding to a constant spacetime field $\phi$.   It is of interest to
obtain the higher-point amplitudes from the field equations (\ref{feom}),
which should be zero if they are to reproduce the string interactions.

    To summarise, assuming (a) the world-sheet field content
$\left(p_\alpha,\, \theta,\, X^{\alpha\dot\alpha}\right)$,  (b)
irreducibility and (c) the quadratic nature of the constraints, we have
shown that there exists one more possible string, in addition to the two
previously studied strings, in 2+2 dimensions. Among the three, only the
string theory constructed in \cite{klpswx} has spacetime supersymmetry.
There are several possible generalisations of this work. For example, one
may introduce extra world-sheet fields. Furthermore, constraints that are
higher order polynomials in basic variables could be considered. Finally, an
interesting open problem concerns the possibility of extending these
constructions to higher-dimensional spacetimes.


\begin{thebibliography}{20}
\frenchspacing


\bibitem{foot} M. Ademollo, L. Brink, A. D'Dadda, R. D'Auria, E. Napolitano,
S. Sciuto, E. Del Giudice, P. Di Vecchia, S. Ferrara, F. Gliozzi, R. Musto,
R. Pettorini and J.H. Schwarz, {\em Nucl. Phys.} {\bf B111} (1976) 77.
\bibitem{ov} H. Ooguri and C. Vafa, {\em Mod. Phys. Lett.} {\bf A5} (1990)
1389; {\em Nucl. Phys.} {\bf B361} (1991) 469.
\bibitem{klpswx} Z. Khviengia, H. L\"u, C.N. Pope, E. Sezgin, X.J. Wang and
K.W. Xu, {\it $N=1$ superstring in 2+2 dimensions}, preprint, CTP TAMU-2/95,
hep-th/9504121.
\bibitem{ws1} W. Siegel, {\em Phys. Rev. Lett.} {\bf 69} (1992) 1493.
\bibitem{lp} H. L\"u and C.N. Pope, {\it BRST quantisation of the $N=2$
string}, CTP TAMU-62/94, SISSA-175/94/EP, hep-th/9411101.

\end{thebibliography}
\end{document}